\documentclass[twocolumn]{webofc}
\usepackage[varg]{txfonts}   
\usepackage[utf8]{inputenc}

\begin{document}

\title{Corrections of alpha- and proton-decay energies in implantation experiments}

\author{\firstname{W.J.} \lastname{Huang}\inst{1}\fnsep\thanks{\email{huang@csnsm.in2p3.fr}}
        \and
        \firstname{G.} \lastname{Audi}\inst{1}
}

\institute{CSNSM, Univ Paris-Sud, CNRS/IN2P3, Universit\'{e} Paris-Saclay, Orsay, France}

\abstract{%
Energies from alpha- and proton-decay experiments yield information of capital importance for deriving the atomic masses of superheavy and exotic nuclides.
We present a procedure to correct the published decay energies in case the recoiling daughter nuclides were not considered
properly in implantation experiments.
A program has been developed based on Lindhard's integral theory,
which can accurately predict the energy deposition of heavy atomic projectiles in matter.

}
\maketitle

\section{Introduction}
\label{intro}
The study of different decay modes reveals important nuclear structure information.
In particularly, $\alpha$ decay and proton decay are two unique tools to explore the most proton-rich
atomic nuclei~\cite{2013alpha, 2008Blank}.
According to the latest Atomic Mass Evaluation (AME)~\cite{2012Audi},
around 65\% of the input data in the mass range $A>200$ result from $\alpha$-decay experiments.
In lighter mass regions
there are a large number of proton-decay data which share many similarities with $\alpha$-decay data.
Energies from $\alpha$ and proton decay yield information of capital importance for deriving mass values.
There are four major experimental approaches for $\alpha$-decay measurements:
The first one uses a magnetic spectrograph~\cite{1971Grennberg},
from which $\alpha$-kinetic energies are determined by direct measurements of the orbit diameters and the magnetic induction field.
All $\alpha$-energy standards use this method.
The second one uses the scintillating bolometer technique which detects the total $\alpha$-decay energy at temperatures below 100~mK~\cite{2003Marcillac}.
In the third method the nuclide of interest is implanted into a foil and the $\alpha$ particle is detected by surrounding Si detectors ~\cite{2010Andreyev}.
Last but not least the radioactive species, which are produced in a nuclear reaction are directly implanted
into a Si detector: e.g. a double-sided silicon-strip detector (DSSD) or a resistive-strip detector~\cite{2010Knoll}.
The first three methods measure either the pure $\alpha$-particle energy or the total $\alpha$-decay energy,
while the implantation method detects the $\alpha$ (or proton) particle and the heavy recoil daughter nuclide in coincidence.
The knowledge of the behaviour of the recoil nuclide is crucial for obtaining the accurate decay-energy value.

\section{Energy calibration}
In the $\alpha$-decay implantation in detector experiments, authors often make the simple assumption that
only the $\alpha$-particle energy is measured in the detector while in the proton decay,
it is often considered that both the proton and the heavy recoil are detected at the same time
but neither of these statements is correct: $\alpha$-particles and protons with energies of a few MeV have almost 100\%
detection efficiency, which is not the case for the heavy species.

\begin{figure}[!htbp]
\centering
\includegraphics[width=8cm]{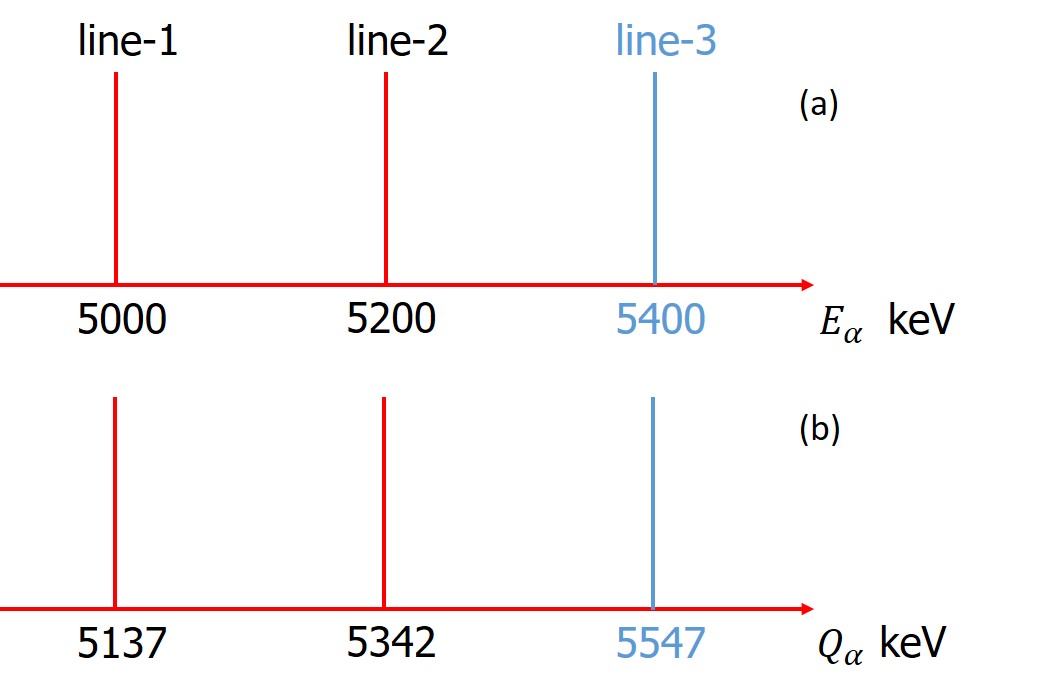}
\caption{Example $\alpha$-decay spectrum where line-1 and line-2 are calibrants and line-3 is unknown.
(a) Case for which the detector detects only the $\alpha$-particle energy. (b) Case where the detector detects also the recoiling nuclide.}
\label{cal}
\end{figure}

Suppose there are three equidistant lines in an $\alpha$-decay spectrum (see Fig.~\ref{cal}).

Two well-known $\alpha$-energy activities line-1 (with $E({\alpha_{1}})=5000$~keV) and line-2 (with $E({\alpha}_{2})=5200$~keV)
are used as calibrants and line-3 is assigned to the unknown nuclide.
If the detector does not detect the recoiling nuclide as in Fig.~\ref{cal}~(a), then what is measured would be the $\alpha$-particle energy
and $E(\alpha_{3})=5400$~keV is easily obtained.
In the other extreme case, when the detector measures all the energy of the recoiling ion, then the energy scale will change as in Fig.~\ref{cal}~(b).
If line-1 and line-2 correspond to a nuclide of mass number $A=150$, the new scales will change to $Q_{\alpha}(\textnormal{line-1})=5137$~keV and $Q_{\alpha}\textnormal{(line-2)}=5342$~keV based on
the simple relation:
\begin{equation} \label{eq:etoQ}
Q_{\alpha} = \frac{M}{M-M_{\textnormal{4He}}}E_{\alpha}
\end{equation}
where $M$ is the mass number of the parent nuclide and $M_{\textnormal{4He}}$ is the mass number of helium-4.
In this case we measure the $\alpha$-decay energy $Q_{\alpha}$ and obtain $Q_{\alpha}(\textnormal{line-3})=5547$~keV.

If line-3 corresponds to a nuclide of mass number $A=150$, its energy $E_{\alpha}$ is deduced to be 5399~keV according to the transformation of Eq.~\ref{eq:etoQ}, which is 1~keV smaller than
the value obtained from Fig.~\ref{cal}~(a).
However, if line-3 corresponds to a nuclide with a different mass number for example, $A=200$, $E_{\alpha}$ will increase from 5400~keV to 5436~keV,
which is already off by 36~keV.
Moreover the detector is not 100\% sensitive to the recoiling nuclide and this more relativistic case will be developed in the next section.

\section{Detection efficiency}
The recoiling ions lose their energies in the Si detector in two ways:
excitation and ionization of the electrons of the atoms (electronic process),
or collision with nuclei of the atoms (nuclear process).
The electronic process produces a signal in the detector, while the nuclear process does not.
Knowledge of both processes is important for implantation $\alpha$-decay and proton-decay experiments
where the heavy recoil is detected simultaneously with the light particle.
In 1963 Lindhard et al.~\cite{1963Lindhard} derived a theory to describe these processes, from which the detection efficiency $K$ was defined as:

\begin{equation} \label{eq:LLS}
K = \frac{\bar{\eta}_{R}}{E_{R}} = \frac{kg(\epsilon)}{1 + kg(\epsilon)}
\end{equation}

where $\bar{\eta}_{R}$ is the part of the recoiling energy that is effectively detected in the detector,
$E_{R}$ is the total recoiling energy,
$\epsilon$ is called the ``dimensionless reduced energy" related to $E_{R}$,
$k$ is a coefficient related to the mass number and the atomic number of the recoil nuclide and the target nuclide,
$g(\epsilon)$ is a semi-empirical function
(for more details please refer to Ref~\cite{1963Lindhard}).
This theory was derived to predict the detected energy of heavy atomic projectiles in matter and the agreement between calculations and experiments data is remarkable~\cite{1975Ratkowski,1982Hofmann}.

\begin{figure}[!htbp]
\centering
\includegraphics[width=9cm]{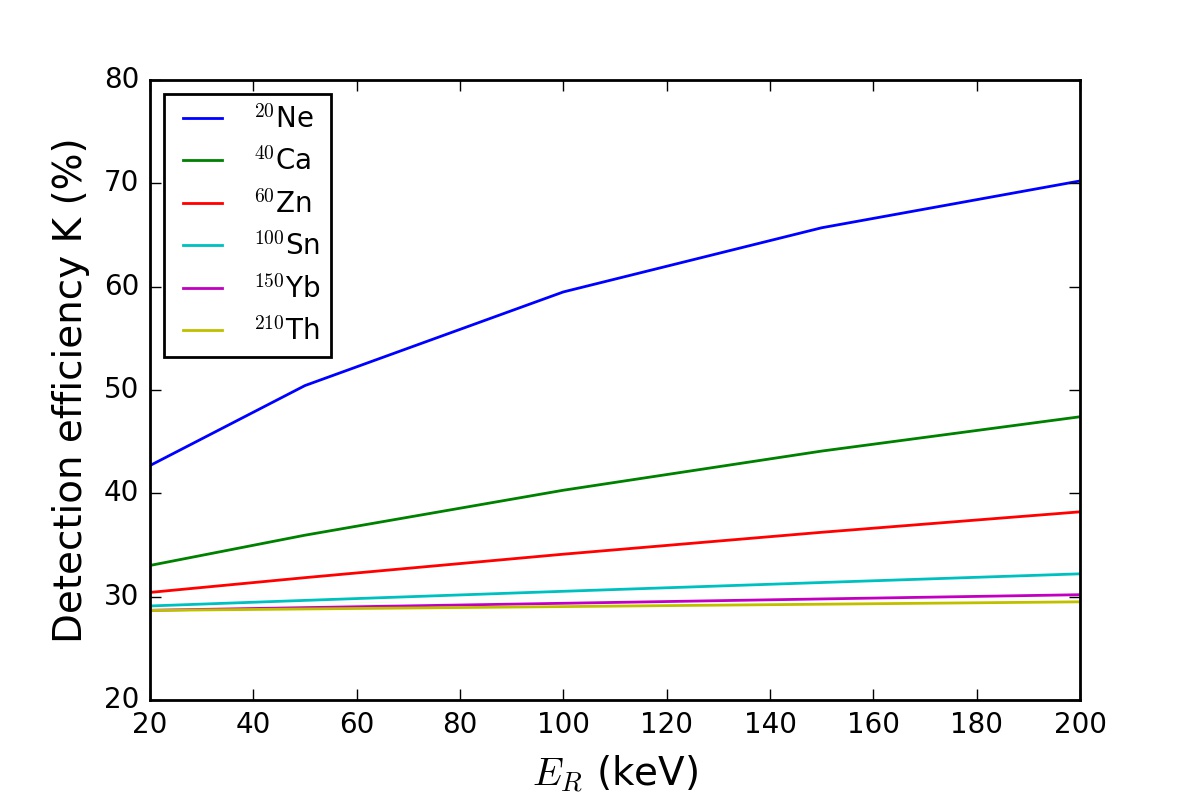}
\caption{Calculation of detection efficiency $K$ for different nuclides at different recoiling energy $E_{R}$ in Si-detector.
The range of $E_{R}$ selected here covers most decay experiment cases.}
\label{K}
\end{figure}

Fig.~\ref{K} shows the calculations of the detection efficiency $K$ for different nuclides based on Lindhard's theory.
For light nuclides (e.g.$^{20}$Ne and $^{40}$Ca), the detection efficiencies increase rapidly as their energies increase.
For intermediate (e.g.$^{60}$Zn and $^{100}$Sn) and heavy nuclides (e.g.$^{150}$Yb and $^{210}$Th), the detection efficiencies increase much more slowly than those of the light nuclides.
For $\alpha$ particles and protons with energies larger than 1~MeV, both detection efficiencies can be considered to be 100\%.
For the implantation method where both the energies of the emitted particles and a part of the heavy recoil are detected, one needs to consider properly the energy loss of the heavy recoil in the detector.
Some experimentalists have already noticed this effect and made the correction for their results~\cite{1991Borrel,1996Blank,2012Hofmann}.
In the following we come up with a concept about how to treat the calibration line and make a correction to the published experimental result,
when the partial recoiling effect was not taken into account.

Here we take $\alpha$ decay as an example.
If we consider the recoiling energy, the new scale should be adjusted to:

\begin{equation} \label{eq:Edetected}
E_{d} = E_{\alpha} + E_{R}*K
\end{equation}

where $E_{d}$ is the total detected energy,
$E_{\alpha}$ is the kinetic energy of the $\alpha$ particle, $E_{R}$ is the recoiling energy and $K$ is the detection efficiency for the recoil nuclide at energy $E_{R}$.
It is $E_{d}$ that should be used in the energy calibration rather than $E_{\alpha}$. Also the recoiling energy can be expressed as:

\begin{equation} \label{eq:Erecoil}
E_{R} = \frac{4}{M-4}E_{\alpha}
\end{equation}
where $M$ is the mass number of the mother nuclide.
Combining Eq.~\ref{eq:Edetected} and Eq.~\ref{eq:Erecoil}, the pure $\alpha$-particle energy can be obtained:

\begin{equation} \label{eq:Epure}
E_{\alpha} = \frac{E_{d}}{1+\frac{4K}{M-4}}
\end{equation}

For proton-decay experiments where $Q_{p}$ is often used in the calibration (as one considers erroneously that the energies of the proton and of the heavy recoil nuclide are fully detected at the same time),
one can obtain a similar relation as Eq.~\ref{eq:Edetected}:

\begin{equation} \label{eq:Qdetected}
E_{d} = E_{p} + E_{R}*K
\end{equation}
where $E_{p}$ is the proton energy and

\begin{equation} \label{eq:Qp1}
E_{p} = \frac{M-1}{M}Q_{p}
\end{equation}

\begin{equation} \label{eq:Qpr}
E_{R} = \frac{1}{M}Q_{p}
\end{equation}
for the proton decay.

Combining Eq.~\ref{eq:Qdetected}, \ref{eq:Qp1} and \ref{eq:Qpr}, one can obtain:

\begin{equation} \label{eq:Qp}
Q_{p}=\frac{M}{M-1+K}E_{d}
\end{equation}

In the next section, we will illustrate how to make the correction for some experimental results.

\section{Application}
\subsection{$^{255}$Lr$^{m}$($\alpha$)}
In Ref~\cite{2008Hauschild}, the detector was calibrated using the well-known $\alpha$-particle energy 7923(4)~keV of $^{216}$Th ~\cite{2016Lopez-Martens}.
The recoiling energy of the daughter nuclide $^{212}$Ra is calculated as $7923*4/212\approx150$~keV and at this energy the detection efficiency $K$ is $29.12\%$.
The calibration line of $^{216}$Th should be adjusted to $E_{d}(^{216}\textnormal{Th})=7923+150*0.2912=7967$ keV.
In the $\alpha$-decay spectrum, the $\alpha$-particle energy of $^{255}$Lr$^{m}$ is 8371 keV,
from which the detected energy of $^{255}$Lr$^{m}$ can be deduced as $E_{d}(^{255}\textnormal{Lr})=7967*8371/7923=8417$ keV.
The recoiling energy of the $\alpha$-decay daughter nuclide $^{251}$Md can be calculated approximately as $8417*4/255\approx131$ keV and at this energy,
its detection efficiency is 29.08\%.
According to the Eq.~\ref{eq:Epure}, the pure $\alpha$-particle energy of $^{255}$Lr$^{m}$ is calculated to be 8378 keV.
The difference between the published value and the corrected value is 7(10) keV.
The same routine can be applied to the $\alpha$-decay energy of the $^{255}$Lr ground state.

\subsection{$^{69}$Kr($\beta$p)}
In Ref~\cite{2014De}, the $\beta$-delayed proton-decay energy of $^{69}$Kr was determined to be 2939(22)~keV using known $\beta$-delayed proton decay energies of 806, 1679, 2692~keV for $^{20}$Mg and 1320, 2400, 2830, 3020, 3650~keV for $^{23}$Si.
The authors assumed (erroneously) that the recoil energy would be fully recorded at the same time~\cite{2015Meisel}.
As one can see from Fig.~\ref{K} the detection efficiency for the intermediate nuclide e.g.$^{60}$Zn, is between 30\%$\sim$40\% and
its neighbouring nuclides show similar behaviour.
The recoiling energy of the $\beta$-delayed proton-decay $^{23}$Si at 3020 keV is $3020/23\approx131$ keV and the detection efficiency for the decay daughter nuclide $^{22}$Mg is 59.75\%.
The effectively detected energy of  this calibration line is 2967 keV according to Eq.~\ref{eq:Qdetected}.
The detected energy of $\beta$-delayed proton-decay nuclide $^{69}$Kr is deduced to be $2967*2939/3020\approx2887$ keV.
The detection efficiency of the daughter nuclide $^{68}$Se is 30.79\% at the corresponding recoiling energy.
Applying Eq.~\ref{eq:Qp}, the $\beta$-delayed proton decay energy of $^{69}$Kr is calculated to be 2916 keV.
The difference between the corrected value and the published one is 23(22) keV, which exceeds 1$\sigma$.\\

From the two examples discussed above,
we demonstrated that the recoiling effect should not be ignored.
In Ref~\cite{2012Hofmann}, the detection efficiency $K$ was assumed to be 0.28 and was applied to all the calibration lines and the nuclide of interest.
It is reasonable to use $K=0.28$ universally in this case as one can see from Fig.~\ref{K} that $K$ becomes almost constant for heavy nuclides.
For light nuclides, $K$ differs quite a lot (59.75\% for $^{22}$Mg and 30.79\% for $^{68}$Se) and should be treated differently.

\section{Conclusion}

As the implantation method is widely used for decay experiments, the effect of the recoil nuclide should be carefully
taken into account.
Lindhard's theory predicts quite well the energy deposition of heavy nuclides in matter and it bas been proven to be reliable by Ref~\cite{1975Ratkowski, 1982Hofmann}.
We propose a way to correct the result if the recoiling effect was not
considered in the energy calibration.
Here we strongly recommend that the authors specify in the publication how they treat the recoil nuclide in the experiment.
Our next step will be to reexamine all the precise alpha- and proton-decay energy data and make the required corrections when necessary.

\section{Acknowledgement}
We are very indebted to Dr.~A.~Lopez-Martens for enlightening discussions,
providing detailed experimental information on Ref~\cite{2008Hauschild}
and careful correcting the paper.
We would like to thank Dr.~M.~Wang for raising this problem and careful checking all the materials.
We thank Dr.~F.~Kondev for the discussion on different alpha-decay methods
and Dr.~D.~Lunney for his wise advices.
This work is supported by China Scholarship Council (CSC No.201404910496).

\end{document}